\documentstyle[12pt]{article}
\title{Four various string baryon models\\
and Regge trajectories}
\author{\large G.S. Sharov \\[3mm]
\em Tver state university, \\
\em 170002, Sadovyj per., 35, Tver, Russia}
\date{}
\begin{document}
\maketitle
\begin{abstract}

Rotational motions for the quark-diquark (q-qq), linear (q-q-q)
``three-string" (Y) and "triangle" string baryon models are considered
and applied to description of baryonic orbitally exited states on
the Regge trajectories. For the model "triangle" (three pointlike
masses bounded pairwise by relativistic strings) the rotating
closed string has a form of a curvilinear triangle composed of
segments of a hypocycloid.

The ultrarelativistic asymptotic relation
between the energy $E=M$ and the total angular momentum
$J$ has the same form
$J\simeq\alpha'E^2+\alpha_1E^{1/2}$
for rotational states in all string baryon models.
Taking into account some type of quark spin-orbit interaction
one can describe the baryonic Regge trajectories by means
of all mentioned string baryon configurations under
assumption of different effective string tensions for them.
Model depending estimations of effective quark
masses in this procedure
$m_{ud}\simeq 130$ MeV, $m_{s}\simeq 270$ MeV
are close for all string configurations.

\end{abstract}

\begin{figure}[b]
\begin{picture}(150,28)
\thicklines
\put(10,15){\line(1,0){25}}
\put(10,15){\circle*{2}}\put(35,16){\circle*{2}}\put(35,14){\circle*{2}}
\put(9,11){q}\put(32,9){qq}\put(12,26){a)}
\put(50,15){\line(1,0){28}}
\put(50,15){\circle*{2}}\put(64,15){\circle*{2}}\put(78,15){\circle*{2}}
\put(49,11){q}\put(63,11){q}\put(77,11){q}\put(52,26){b)}
\put(98,15){\line(0,-1){12}}\put(98,15){\line(-2,1){10}}
\put(98,15){\line(2,1){10}}
\put(98,3){\circle*{2}}\put(88,20){\circle*{2}}\put(108,20){\circle*{2}}
\put(94,3){q}\put(85,17){q}\put(109,17){q}\put(93,26){c)}
\put(115,5){\line(1,0){24}}\put(115,5){\line(2,3){12}}
\put(127,23){\line(2,-3){12}}
\put(115,5){\circle*{2}}\put(139,5){\circle*{2}}\put(127,23){\circle*{2}}
\put(110,5){q}\put(141,5){q}\put(123,25){q}\put(132,26){d)}
\put(62,1){Fig. 1}
\end{picture}
\end{figure}

Shortly after the model of relativistic string with massive ends
describing the meson Regge trajectories [1,2] string models
of baryon were suggested in some variants [3--7]:
a) the quark-diquark model, b) the linear configuration,
c) the ``three-string" model or Y-configuration with three
strings joined in the fourth massless point (junction),
and d) the ``triangle" model or $\Delta$-configuration that could
be regarded as a closed string carrying three pointlike masses (Fig.~1).
Two of these models --- the Y-configuration [3,4] and meson-like
quark-diquark model [5] were investigated in a more detailed way
in comparison with two others. The system q-q-q (b)
seems to be unstable with respect to transformation to the q-qq
configuration (a) [5].

The investigation of the ``triangle" model (d) encounter some
difficulties, which were overcome in Ref.~[6] for
a set of classic uniform planar rotations of the system.

All mentioned string models could be constructed
on the base of the action
\begin{equation}
S=-\int\limits_{\tau_1}^{\tau_2}\!\biggl\{\gamma\!
\int\limits_{\sigma_*(\tau)}^{\sigma_N(\tau)}\!\!
\sqrt{(\dot XX')^2-\dot X^2X'{}^2}\,d\sigma+
\sum_{i=1}^Nm_i\sqrt{V_i^2(\tau)}\biggr\}d\tau.
\end{equation}
Here $\gamma$ is the string tension, $x^\mu=X^\mu(\tau,\sigma)$ ---
the world surface, $m_i$ --- quark (or diquark) masses,
$\,\dot X^\mu=\partial_\tau X^\mu$,
$X'{}^\mu=\partial_\sigma X^\mu$, $c=1$, $\hbar=1$,
$V_i^\mu=\frac d{d\tau}X^\mu(\tau,\sigma_i(\tau))$ is a tangent
vector to the i-th quark trajectory $\sigma=\sigma_i(\tau)$.
The number $N=2$, $\sigma_*=\sigma_1(\tau)$ for the quark-diquark
model and $N=3$ for other models of baryon.
In the linear configuration q-q-q $\sigma_*=\sigma_1(\tau)$,
$\sigma_i(\tau)<\sigma_{i+1}(\tau)$.

In the ``triangle" model the equations $\sigma=\sigma_*=\sigma_0(\tau)$
and $\sigma=\sigma_3(\tau)$ define the trajectory of the same ---
3-rd quark. It may be written as the closure
condition $X^\mu(\tau,\sigma_0(\tau))=X^\mu(\tau^*,\sigma_3(\tau^*))$.
The parameters $\tau$ and $\tau^*$ aren't equal in general.

The equations of motion and the boundary conditions on the quark
trajectories in these models are deduced by variation and minimization
of action (1).

Under the conditions of orthonormality $\dot X^2+X'{}^2=0$,
$(\dot XX')=0$ (they may be obtained for all configurations [2,6])
the equations of motion become linear
\begin{equation}\ddot X^\mu-X''{}^\mu=0,\end{equation}
and the boundary conditions take the simplest form.
In particular, for the ``triangle" model and for the middle
quark in the q-q-q configuration this form is
\equation
m_i\frac d{d\tau}\frac{V_i^\mu}{\sqrt{V_i^2}}
-\gamma\bigl(X'{}^\mu+\sigma_i'(\tau)\,\dot X^\mu\bigr)
\Big|_{\sigma=\sigma_i+0}\!\!
+\gamma\bigl(X'{}^\mu+\sigma_i'(\tau)\,\dot X^\mu\bigr)
\Big|_{\sigma=\sigma_i-0}\!\!=0.
\endequation
In these models $\dot X^\mu$ and $X'{}^\mu$ may have
discontinuities on the lines $\sigma=\sigma_i(\tau)$.

The solution of Eq.~(2) satisfying the above conditions and describing
the uniform rotation of the rectilinear string is well known for the meson
string model [1,2] and for the q-qq ($N=2$) and q-q-q ($N=3$)
configurations can be represented as
\equation
X^\mu=\bigl\{\tau;\,\Omega^{-1}\sin\Omega\sigma\cdot\cos\Omega\tau;
\,\Omega^{-1}\sin\Omega\sigma\cdot\sin\Omega\tau\bigr\}.\endequation
Here $\Omega$ is the angular velocity, $\sigma\in[\sigma_1,\sigma_N]$,
$\sigma_i={}$const, $\sigma_1<0$, $\sigma_N>0$.

The rotational motion of the ``three-string" model with the junction at rest
and with rectilinear string segments, forming in a plane of rotation angles
120${}^\circ$ [3,4], is described by the expression similar to Eq.~(4).

For the q-qq, q-q-q and Y configurations the energy $E=M$ and the
angular momentum $J$ ($z$ projection) of
the considered rotational motions are [2\,--\,5]
\equation
\!E=\sum_{i=1}^N\bigg[\frac\gamma\Omega\arcsin v_i+
\frac{m_i}{\sqrt{1-v_i^2}}\bigg]+\Delta E,\;\;
J=\frac1{2\Omega}\sum_{i=1}^N\bigg(\frac\gamma\Omega
\arcsin v_i+\frac{m_iv_i^2}{\sqrt{1-v_i^2}}\bigg)+S,
\endequation
where the velocities of moving quarks
$v_i=\sin|\Omega\sigma_i|=\sqrt{\Big(\frac{m_i\Omega}{2\gamma}\Big)^2+1}-
\frac{m_i\Omega}{2\gamma}$.
The presence of quark spins with projections $s_i$ ($\sum_{i=1}^Ns_i=S$)
is taken into account as a small correction to the classic motion
and to the energy $\Delta E\simeq\Delta E_{SL}$
(the spin-spin correction is assumed to be small in comparison with the
spin-orbit one at high $J$).

The rotational motions (uniform rotations about the system center of mass)
in the ``triangle" model [6] may be presented in the form
\equation
X^0\equiv t=a\tau+b\sigma,\qquad
X^1={}\mbox{Re}\,u(\sigma)\,e^{i\omega\tau},\quad
X^2={}\mbox{Im}\,u(\sigma)\,e^{i\omega\tau}.
\endequation
Here the complex function
$u(\sigma)=A_i\cos\omega\sigma+B_i\sin\omega\sigma$,
$\sigma\in[\sigma_i,\sigma_{i+1}]$ is continuous in
$[\sigma_0,\sigma_3]$, the values $\sigma_i$,
$\dot X^2(\tau,\sigma_i)=V_i^2$, $\tau^*-\tau=T$
are the constants. Owing to Eqs.~(3) six complex constants $A_i,B_i$
are proportional to $A_1$ [6,7]
\begin{eqnarray*}
&A_0=A_1,\quad B_0=\textstyle\frac12(h_1-iK)\,A_1,
\quad B_1=-\frac12(h_1+iK)\,A_1,&\\
&A_2=\big[1+\frac12h_2d_1(2c_1-h_1d_1-iKd_1)\big]\,A_1,&\\
&B_2=-\frac12\big[2h_2c_1^2+(1-h_2c_1d_1)(h_1-iK)\big]\,A_1,&
\end{eqnarray*}
and real $a$ and $b$ --- to $|A_1|$:
$a=\sqrt{D\omega^2K|A_1|^2/(2T)}$, $b=-aT/D$, where
$D=\sigma_3-\sigma_0$, $K=2d_2d_0^{-1}d_1^{-1}\sin\omega T/(G_2G_3-1)$,
$d_i=\sin\omega(\sigma_{i+1}-\sigma_i)$,
$c_i=\cos\omega(\sigma_{i+1}-\sigma_i)$,
$G_i=(h_id_{i-1}d_i-d_{i-1}c_i-c_{i-1}d_i)/d_{i+1}$,
$h_i=\omega m_i\gamma^{-1}|V_i|^{-1}$. The notations
$c_i,\,d_i,\,h_i,\,G_i$ are cyclical: $á_{i+3}\equiv á_i$,
$d_{i+3}\equiv d_i\dots$

Expression (6) is the solution of Eq.(2) and satisfies the
orthonormality, closure and boundary (3) conditions if the
parameters are connected by the relations [6,7]
\begin{eqnarray*}
&2\cos\omega T=G_1+G_2+G_3-G_1G_2G_3,\quad
D/T+T/D=(K^2+4+h_1^2)/(2K),&\\
&(G_{i+1}-G_i)\,d_i=(G_iG_{i+1}-1)(d_{i-1}c_{i+1}-c_{i-1}d_{i+1}).&
\end{eqnarray*}
For given $m_i$, $\gamma$ and a parameter measuring the rotational speed
one can find all mentioned values by solving these equations.

A string position for this state (a section $t={}$const of the
surface (6)) is the curve composed of three segments of a hypocycloid
joined in three points (the quark positions).
Hypocycloid is the curve drawing by a point of a circle that is rolling
in another fixed circle with larger radius. Here $R/r=2/(1-|b|/a)$.

The energy and the angular momentum of this state are [6]
\equation
\!E=\gamma D\frac{a^2-b^2}a+\sum_{i=1}^3\frac{m_i}{\sqrt{1-v_i^2}}+
\Delta E,\;\;
J=\frac1{2\Omega}\bigg(\gamma D\frac{a^2-b^2}a
+\sum_{i=1}^3\frac{m_iv_i^2}{\sqrt{1-v_i^2}}\bigg)+S.
\endequation
where the quark velocities
$v_i=\sqrt{Td_{i-1}d_i(G_{i-1}G_{i+1}-1)\big/(Dd_{i+1}\sin\omega T)}$,
$\Omega=\omega/a$.

A set of topologically different configurations of the system
classified [7] with using the fact
that in the limit $v_i\to1$ the string configuration tends to
the hypocycloid described by integer
$n=\lim\limits_{v_i\to1}\frac{|\omega|D}\pi$ and
$k=\lim\limits_{v_i\to1}\frac{\omega T}\pi$. Here $n$ is number of the
cusps, $k=n-2,\,n-4,\dots,-n+2$, $R/r=2/(1-|k|/n)$.
In the case $n=2$, $k=0$ (``quark-diquark" state) two quarks coincide,
one of $d_i$ equals 0, $T=b=0$. In this state the world surface (6)
reduces to (4) and the model ``triangle" --- to the q-qq one with the
tension $2\gamma$ (rectilinear segment is the particular case of the
hypocycloid with $R/r=2$).
In the state $n=3$, $k=1$ (simple state) the string has a triangle form,
for other $n$ and $k$ (exotic states) --- more
complicated form with the cusps moving at the speed of light [7].
The simple states of this model
are used below for describing the Regge trajectories.

Two different forms of the spin-orbit correction to the energy are [5,8]
\begin{eqnarray}
\textstyle \Delta E_{SL}=\sum\limits_i\beta(v_i)(\Omega s_i),&
\;\mbox{where}\;&\beta(v_i)=-\big[(1-v_i^2)^{-1/2}-1\big],\\
&\mbox{or}&\beta(v_i)= 1-(1-v_i^2)^{1/2}.
\end{eqnarray}
The expression (8) is due to the Thomas precession of the quark spins.
It is obtained under the assumption that in the quark rest frame
the field is pure chromoelectric [5,8]. The alternative assumption
about pure chromoelectric field in the rotational center rest frame
results in Eq.~(9).

The ultrarelativistic asymptotic behaviour of the dependence $J(E)$
(5) and (7) for q-qq, Y and $\Delta$-configurations has the form
\begin{equation}
J\simeq\alpha'E^2-\nu E^{1/2}\sum_{i=1}^Nm_i^{3/2}
+\sum_{i=1}^Ns_i\big[1-\beta(v_i)\big],
\quad v_i\to1,\end{equation}
where $\displaystyle\alpha'=\frac1{2\pi\gamma}\cdot\left\{
\begin{array}{lr}1, &\mbox{q-qq},\\
2/3, &\mbox{Y},\\  n/(n^2-k^2), &\Delta,\end{array}\right.\quad
\nu=\frac1{\sqrt\pi\,\gamma}\cdot\left\{
\begin{array}{lr}2/3, &\mbox{q-qq},\\
(2/3)^{3/2}, &\mbox{Y},\\
\frac{\sqrt2\,n}3(n^2-k^2)^{-3/4}, &\Delta.
\end{array}\right.$
The case q-q-q differ from q-qq by the substitution $E\;\to\;E-m_2$
in Eq.~(10).

The Regge slope $\alpha'\simeq0.9$  GeV${}^{-2}$ is close
for mesons and baryons. So the effective value of string tension $\gamma$
in the models Y and ``triangle" (simple states) is to be
\equation
\gamma_Y=\frac23\gamma,\quad\gamma_\Delta=\frac38\gamma,\quad
\gamma=\gamma_{q-qq}=\gamma_{q-q-q}\simeq\frac{1.1}{2\pi}\,\,(\mbox{GeV}^2).
\endequation
It is probably connected with different
energies of QCD interaction in the pairs: quark-quark and
quark-antiquark.

The criterion for choosing the form of spin-orbit correction (8) or (9)
results from applying the relation (10) to two Regge
trajectories: nucleonic $N_\alpha$-trajectory with $S=1/2$ and
$\Delta$-trajectory with $S=3/2$ (Fig.~2). In accordance with
Eq.~(10) the latter is to be higher by $\Delta J=1-\beta(v_i)$.
This comparison compels us to choose expression (9).
The opposite choice (Thomas type precession) is possible only for q-qq
model under assumption that the diquark mass is different for the
mentioned trajectories (220 and 550 MeV in Ref.~[5]).

For given $\beta(v_i)$ (9) one can estimate the effective quark masses
with using Eq.~(10). These estimations for the N, $\Delta$
and $\Lambda$ baryons give the values $m_{ud}\simeq 130$ MeV,
$m_{s}\simeq 270$ MeV, which under assumptions (11)
are close for q-qq, Y and ``triangle" string configurations.
As to a choice of the configuration ---
the q-qq one is preferable, if we assume the principle of minimal
energy for the string system with given $J$ [5].
The QSD Wilson loop operator approach gives
some arguments in favour of the Y [9] or ``triangle" model [10].

The dependence $J(E)$ (5) and (7) for these models for some N,
$\Delta$ baryons is shown in Fig.~2 and for strange baryons ---
in Fig.~3.

The chosen values $m_{ud}$, $m_{s}$, $\beta(v_i)$ result in
satisfactory description of the Regge trajectories for these
baryons and also for light and strange mesons.

The string solutions obtained in the ``triangle" model (they also take
place for a closed massless string that has a form of a rotating
hypocycloid with the cusps moving at the speed of light) are probably
applicable not only to the particle physics.

\medskip

\centerline{\bf References}

\smallskip

\noindent[1] {\it Y. Nambu} // Phys. Rev. D. {\bf 10}, 4262 (1971);
{\it A. Chodos, C.B. Thorn}, Nucl. phys. {\bf B72}, 509 (1974).
\newline[2] {\it B.M. Barbashov, V.V. Nesterenko}, Introduction to the
relativistic string theory. Singapore: World scientific, 1990.
\newline[3] {\it X. Artru}, Nucl. phys. {\bf B85}, 442 (1975);
{\it P.A. Collins, J.F.L. Hopkinson}, ibid. {\bf B100}, 157
(1975); {\it K. Sundermeyer, A. de la Torre}, Phys.Rev.D {\bf15},
1745 (1977).
\newline[4] {\it M.S. Plyuschai, G.P. Pron'ko, A.V. Razumov},
Theor. Mathem. Physics. {\bf 63}, 389 (1985); {\it S.V. Klimenko et al.},
Theor. Mathem. Physics. {\bf 64}, 810 (1985).
\newline[5] {\it Yu.I. Kobzarev, L.A. Kondratyuk, B.V. Martemyanov,
M.G. Schepkin,} Yad. Fiz. {\bf 45}, 526 (1987).
\newline[6] {\it G.S. Sharov,} Theor. Mathem. Physics. {\bf 113}, 1263 (1997).
\newline[7] {\it G.S. Sharov,} Theor. Mathem. Physics. {\bf 114}, 277 (1998).
\newline[8] {\it T.J. Allen, M.C. Olsson, S. Veseli, K. Williams,}
 Phys. Rev. D {\bf 55}, 5408 (1997).
\newline[9] {\it N. Isgur, J. Paton,} Phys. Rev. D {\bf 31}, 2910 (1985);
{\it Yu.S. Kalashnikova, A.V. Nefediev,} Yad. Fiz. {\bf 60}, 1470 (1997).
\newline[10] {\it J. M. Cornwall,} Phys. Rev. D {\bf 54}, 6527 (1996).

\end{document}